\documentclass[cameraready]{Interspeech} 

\author[affiliation=1]{Tristan}{Tsoi}
\author[affiliation=1,correspondingauthor=true]{Jiajun}{Deng}
\author[affiliation=1]{Yingke}{Zhu}
\author[affiliation=2]{Huu Quyen}{Dang}
\author[affiliation=1]{Tianxiang}{Cao}
\author[affiliation=3]{Nikita}{Kuzmin}
\author[affiliation={1,2}]{Tao}{Zhong}
\author[affiliation=1]{Simon}{Lui}

\address{
  $^{1}$ Central Media Technology Institute, Huawei \quad
  $^{2}$ The Chinese University of Hong Kong \\
  $^{3}$ Nanyang Technological University \quad
}

\email{tsoi.tik.nang1@huawei.com, jjdeng321@gmail.com}

\usepackage{booktabs}
\usepackage{multirow}
\usepackage{graphicx}
\usepackage{amsmath}
\usepackage[table]{xcolor}
\usepackage{pifont}

\title{Next-Turn: Duration-Aware Streaming Endpoint Detection via Time-to-Next-Speech-Onset Prediction}

\begin{document}

\maketitle

\begin{abstract}
Endpoint detection (EPD) is essential for natural turn-taking in streaming speech systems. However, reliably determining the endpoint of an utterance is challenging because speakers often pause mid-utterance due to hesitations and disfluencies. Semantic EPD has emerged as a promising direction to address this issue but is hindered by ambiguous supervision and strict streaming constraints. We propose Next-Turn which uses the time-to-next-speech-onset as the training objective, where targets are derived directly from speech timestamps and require no manual additional annotation. Experiments show that the proposed method outperforms conventional acoustic and recent semantic EPD baselines, achieving a 25.9\% absolute improvement in endpoint accuracy within 320 ms over the strongest baseline. In addition, joint training with the duration-aware objective complements standard binary EPD, with gains that increase monotonically with increasing pauses.
\end{abstract}


\vspace{-2mm}
\section{Introduction}
\label{sec:intro}

The growing demand for highly interactive speech interfaces has driven widespread adoption of streaming speech models, particularly for full-duplex conversation \cite{liao2025flexduo, chen2025fireredchat, defossez2024moshi} and real-time speech translation \cite{barrault2023seamless, ren2020simulspeech}. In these settings, the ability to determine when a user has finished a meaningful speech unit is critical for both perceived responsiveness and downstream model performance. Endpoint detection (EPD), also referred to as end-of-utterance or end-of-turn detection, therefore remains a fundamental component of streaming speech processing pipelines.

Conventional EPD methods primarily rely on acoustic cues, typically implemented via neural voice activity detection (VAD) \cite{hughes2013vad, zhang2013dbnvad, ryant13_interspeech} or statistical models \cite{ferrer2002endofutterance, sohn1999vad} that predict speech/non-speech boundaries. Although effective in detecting silence and coarse activity changes, such approaches often conflate acoustic completion with semantic completion. In natural speech, pauses frequently occur mid-thought due to hesitations, disfluencies, or planning, and thus silence alone is an unreliable indicator of end-of-utterance \cite{heldner2010pauses}. Consequently, acoustic-only endpoint decision may trigger prematurely, truncating a speaker before the intended message is complete, or trigger too late, introducing unnecessary tail latency \cite{li2020faststreamingasr, huang2022e2esegmenter}. These limitations are particularly harmful in full-duplex spoken interaction and real-time speech translation.

To mitigate this mismatch, a line of work incorporates linguistic or semantic evidence by leveraging automatic speech recognition (ASR) hypotheses and performing text-based endpoint prediction from transcripts \cite{chen2025fireredchat, wang2024turntaking, zhang2025llmdm, bijwadia2022unified}. Although ASR-conditioned methods provide richer semantic cues, they also make endpoint decisions sensitive to ASR errors and instability and couple both runtime cost and decision latency to the cascade pipeline \cite{wu2025phoenixvad}. Motivated by these constraints, recent work has explored audio-only semantic EPD models that learn representations from the waveform which capture not only speech presence but also cues of semantic completion, thereby retaining the low-overhead, plug-and-play properties of VAD-like components while improving semantic awareness \cite{liao2025flexduo, wu2025phoenixvad, li2025easyturn, shi2023semanticvad}.

Efforts to develop a single, unified semantic EPD model for real-time speech applications face two fundamental challenges: (i) the lack of reliable supervision for semantic completion, and (ii) the inherent tension between semantic modeling and streaming constraints. First, semantic endpoints are substantially more ambiguous than speech/non-speech boundaries. Whereas speech activity can be annotated with relatively clear acoustic criteria, “completion” depends on discourse context and speaker behavior: hesitations, disfluencies, mid-utterance pauses, and backchannels can all produce silence patterns that resemble true endpoints \cite{heldner2010pauses, raju2024twopassep, liu15d_interspeech}. This makes it difficult to obtain consistent semantic endpoint labels and can lead to noisy training targets that hinder generalization. Second, semantic EPD must operate under low-latency requirements with limited right context for practical streaming applications. Accurately detecting semantic completion typically benefits from longer-range context and higher-level linguistic information, yet streaming EPD must make timely decisions and remain robust when partial evidence later changes (e.g., the speaker continues after a short pause). Designing a model that is both semantically informed and stable in real time therefore remains non-trivial.

To this end, this paper proposes Next-Turn, a duration-aware streaming EPD framework that addresses both supervision ambiguity and real-time semantic reasoning. Motivated by evidence that listeners continuously project the timing of upcoming turn boundaries \cite{deruiter2006projecting, ekstedt2022vap}, we present a duration-aware training objective that uses the time-to-next-speech-onset as the supervision target, which serves as a proxy for the semantic endpoint. The model predicts, at each audio frame, the remaining time until the next speech onset, yielding a continuous confidence measure of endpoint likelihood. The targets are derived directly from speech timestamps and require no additional manual annotation. Additionally, to ensure streaming-compatible semantic modeling, we apply random audio cut-off training to the pretrained semantic audio encoder, randomly truncating future context so the model learns to predict completion cues with limited look-ahead. The main contributions are summarized as follows:

\textbf{1)} This paper presents a unified semantic EPD framework that is streaming, low-latency, and effective. It enables online chunk-level endpoint decisions with limited look-ahead and low computational overhead, making it suitable for real-time deployment.

\textbf{2)} Experiments demonstrate that the proposed method consistently outperforms conventional acoustic EPD and recent state-of-the-art semantic EPD baselines in both endpoint accuracy and detection latency, achieving a 25.9\% absolute improvement in endpoint accuracy within 320 ms over the strongest baseline. Furthermore, joint training with the duration-aware objective complements standard EPD, with gains that increase monotonically with increasing pauses.

\vspace{-2mm}
\section{Binary Endpoint Detection}
\label{sec:binary}
We first describe the binary semantic EPD, which serves both as a baseline and as the foundation for the duration-aware extension in Section~\ref{sec:duration}. 

\subsection{Binary Formulation} 
\label{sec:binary_formulation}
Binary EPD aims to determine whether an utterance has reached endpoint by time $t$. Let \(t_{\text{end}}\) denote the time when the speech ends. We define the binary target \(y(t)\) as:

\begin{equation}
  y(t) = \begin{cases}
    0, & t < t_\text{end}; \\
    1, & t \geq t_\text{end}.
  \end{cases}
  \label{eq:bin_target}
\end{equation}
where \(y(t)=1\) indicates EP and \(y(t)=0\) indicates non-EP. The model is optimized with cross-entropy loss at a single decision time for each utterance. During inference, the model outputs a scalar endpoint score \(s_{\text{bin}}(t) \in [0,1]\).

\subsection{Model Architecture} 
\begin{figure}[t]
  \centering
  \includegraphics[width=0.85\columnwidth]{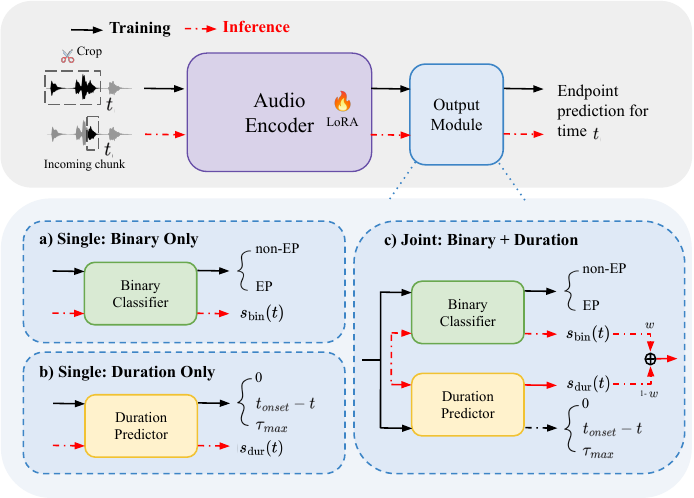}
  \caption{Overview of the proposed Next-Turn framework. a) Single-task with a binary head. b) Single-task with a duration head. c) Joint with a shared encoder and task-specific heads. At inference, the system can use the binary score, the duration-derived score, or a fusion of both (Sec.~\ref{sec:inference_fusion}).}
  \label{fig:architecture}
\end{figure}

Fig.~\ref{fig:architecture} illustrates the overall architecture of our system. We use the Whisper \cite{radford2023whisper} encoder as the audio encoder and fine-tune it with Low-Rank Adaptation (LoRA) \cite{hu2022lora} for parameter-efficient adaptation. The encoder outputs a sequence of frame-level hidden states which we apply mean pooling over the temporal dimension to obtain a single utterance-level representation. A binary classification head (Fig.~\ref{fig:architecture}a), implemented as a linear projection $\mathrm{Linear}(H,2)$ with softmax, maps this pooled representation to EP/non-EP probability, where $H$ is the encoder hidden-state dimension. 

\vspace{-2mm}
\section{Duration-Aware Endpoint Detection}
\label{sec:duration}

We propose a semantic EPD framework based on time-to-next-onset prediction, which provides graded temporal supervision. The duration prediction objective can be used either as an alternative to the binary formulation or jointly with it during training. All duration-aware variants use the same architecture as the binary baseline and differ only in the prediction head(s).

\begin{figure}[h]
  \centering
  \includegraphics[width=0.7\columnwidth]{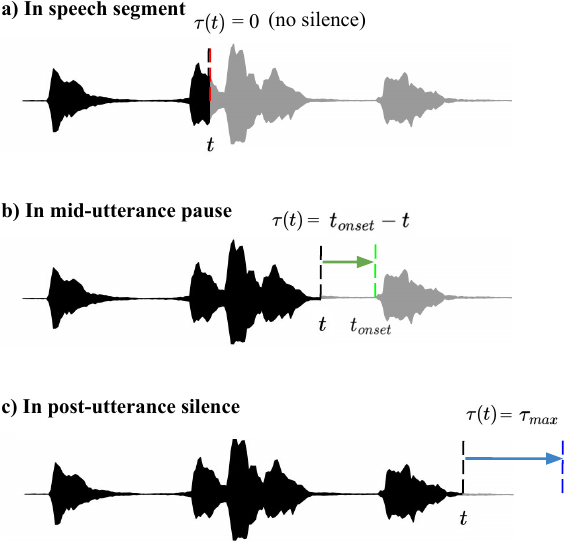}
  \caption{Illustration of the duration target $\tau(t)$ across three regions: a) during speech, $\tau(t) = 0$; b) during mid-utterance pause, $\tau(t)$ is the silence until the next speech onset; and c) after the utterance ends, $\tau(t)$ is set to $\tau_{max}$.}
  \label{fig:formulation}
\end{figure}

\subsection{Duration Formulation}
\label{sec:duration_formulation}
Fig.~\ref{fig:formulation} illustrates the duration target \(\tau(t)\) across three regions: a) in speech segment, b) in mid-utterance pause, and c) in post-utterance silence. At time \(t\), we define \(\tau(t)\) as the remaining time until the next speech onset:

\vspace{-2mm}

\begin{equation}
  \tau(t)=
  \begin{cases}
    0, & t \text{ is in a speech segment};\\
    t_{\text{onset}}-t, & t \text{ is in a mid-utterance pause};\\
    \tau_{\max}, & t \text{ is in post-utterance silence}.
  \end{cases}
  \label{eq:dur_target}
\end{equation}
where $t_{\text{onset}}$ denotes the start time of the next speech segment and $\tau_{\max}$ is a constant assigned to post-utterance, since the true duration until the next turn is unknown within a single utterance. 

\subsection{Duration Prediction Head}

We implement the duration head in two modes (Fig.~\ref{fig:architecture}b): 1) regression (REG) directly regresses $\tau(t)$ with mean-squared-error loss; 2) classification (CLS) which discretizes $\tau(t)$ into $K$ duration classes. The duration head is a two-layer MLP with hidden size $H/2$, ReLU \cite{nair2010relu} activation, and dropout \cite{srivastava2014dropout}, outputting either a scalar (REG) or $K$ logits (CLS), which are normalized with softmax to obtain the class prediction.

During inference, the model predicts a duration value $\hat{\tau}(t)$. This prediction is converted into an endpoint score $s_{\text{dur}}(t)$ by normalizing with $\tau_{\max}$ and clipping to $[0,1]$, i.e., $s_{\text{dur}}(t)=\min(\hat{\tau}(t)/\tau_{\max},\,1)$. For the REG variant, $\hat{\tau}(t)$ corresponds to the regressed duration value, while for the CLS variant, $\hat{\tau}(t)$ is set to the upper boundary of the predicted duration class. 

\subsection{Multi-Task Learning}
The binary objective provides only coarse supervision, while duration prediction supplies a fine-grained temporal signal that complements the binary task. We therefore adopt a multi-task setup that jointly trains binary endpoint classification and duration prediction with a shared encoder (Fig.~\ref{fig:architecture}c). Both heads are optimized using the combined objective $\mathcal{L}=\mathcal{L}_{\text{bin}}+\mathcal{L}_{\text{dur}}$.

\subsection{Score Fusion} 
\label{sec:inference_fusion}
Since the jointly trained binary and duration heads capture complementary views of endpoint likelihood, we optionally combine the two scores from the joint-training system via weighted averaging to obtain a fused endpoint score $\bar{s}(t)$:
\begin{equation}
  \bar{s}(t) = w\,s_\text{bin}(t) + (1{-}w)\,s_\text{dur}(t),
  \label{eq:fusion}
\end{equation}
where $w \in [0, 1]$ (with $w{=}1$ and $w{=}0$ recovering binary-only and duration-only inference, respectively).

\section{Experiments}
\label{sec:experiments}
\subsection{Dataset}
We train on an in-house corpus of 1{,}177 hours of Chinese speech (1{,}097{,}898 utterances), spanning conversational, command-style, and question--answer scenarios, at 16\,kHz. The training set shows a long‑tailed distribution over pause counts. For evaluation, we hold out 1{,}185 utterances disjoint from the training data and manually label their endpoints. The evaluation set is deliberately balanced across pause counts, with 22\%, 18\%, 17\%, 19\%, and 24\% of utterances containing zero, one, two, three, and four or more pauses, respectively. Because annotation is manual, the evaluation set is necessarily modest in size, and absolute results may be sensitive to sampling.

\subsection{Experimental Setup} 
\noindent \textbf{Experimental Setting:} All models use the Whisper-large-v3 \cite{radford2023whisper} encoder unless otherwise stated. LoRA matrices with rank $r{=}8$, scaling factor $\alpha{=}32$, and dropout $p{=}0.05$ are inserted into the query, key, and value projections of every encoder block. Models are optimized with AdamW ($\beta_1{=}0.9$, $\beta_2{=}0.98$, weight decay $0.1$) using a learning rate of $1{\times}10^{-4}$, batch size of 32 across 8 GPUs with 4-step gradient accumulation, bf16 mixed precision, and gradient clipping at 1.0. For the duration formulation, we set $\tau_{\max}=2.0\,\mathrm{s}$. In the CLS variant, $\tau(t)$ is discretized into $K{=}7$ classes corresponding to an in-speech bin, and silence-duration bins $[0,60)$, $[60,120)$, $[120,480)$, $[480,640)$, $[640,800)$, and $[800,\infty)$ ms. The bins are chosen so that each contains approximately the same number of training pause samples. Dropout in the duration head is set to 0.1. We validate every 500 steps and apply early stopping with patience 5, for up to 50{,}000 steps or 10 epochs, whichever is reached first. 

\noindent \textbf{Evaluation Pipeline:} We evaluate all systems in a streaming setting with non-overlapping 160\,ms chunks. To reduce jitter, we smooth scores over neighboring chunks using an exponential-decay weighted average over \(P\) past and \(F\) future chunks:
\begin{equation}
  \bar{s}(t) = \frac{\sum_{j=t-P}^{t+F} \gamma^{|j-t|}\, s(j)}{\sum_{j=t-P}^{t+F} \gamma^{|j-t|}},
\label{eq:smooth}
\end{equation}
\noindent where $\gamma=0.5$. Unless stated otherwise, we use \(P=1\) and \(F=0\). During evaluation, a decision threshold $\theta$ is applied to convert endpoint scores into binary outcomes.

\noindent \textbf{Data Processing:} We first segment each recording into semantically complete utterances
using TEN Turn Detection \cite{ten_turn_detection} on the corpus's reference transcripts. To then obtain speech and silence regions for constructing $\tau(t)$, we derive phoneme-level timestamps via forced alignment using a Kaldi \cite{povey2011kaldi} model trained on the same training data, and produce segment boundaries using a 150 ms silence threshold. During training, each sample begins at the utterance onset and is truncated at a point drawn uniformly from one of three regions (Fig.~\ref{fig:formulation}): a) in a speech segment, b) in a mid-utterance pause, or c) in a post-utterance silence, ensuring exposure to all three conditions in Eq.~\ref{eq:dur_target}. Pause-region truncations are oversampled to mitigate the imbalance between speech-active and pause frames.

\subsection{Evaluation Metrics} 
We use two complementary metrics. For each utterance \(i \in \{1,\dots,N\}\), let
\(t^{*}_{\mathrm{end},i}\) denote the ground-truth endpoint of speech, and let \(\hat{t}_i\) denote the time when the system first triggers an endpoint decision. Here, \(N\) is the total number of evaluated utterances, and \(\delta\) is the tolerated latency window (ms).

\noindent\textbf{Early Interruption ($\mathrm{EI}$)} is the percentage of utterances where the endpoint prediction occurs strictly before $t_\text{end}^{*}$:
\begin{equation}
  \mathrm{EI} = \frac{|\{i : \hat{t}_i < t_{\text{end},i}^{\text{*}}\}|}{N}.
  \label{eq:ei}
\end{equation}
\textbf{Accuracy (\( \mathrm{ACC}_{\delta} \))} is the percentage of utterances where the first EP prediction
occurs within $\delta$ after $t_\text{end}^{*}$:
\begin{equation}
  \mathrm{ACC}_\delta = \frac{|\{i : t_{\text{end},i}^{\text{*}} \le \hat{t}_i \le t_{\text{end},i}^{\text{*}} + \delta\}|}{N}.
  \label{eq:acc}
\end{equation}
We report $\mathrm{ACC}$ at $\delta \in \{160, 320, 480, 640\}$\,ms and optimize for $\mathrm{ACC}_{320}$, as 320\,ms is a standard low-latency target for semantic EPD [15, 19]. $\mathrm{EI}$ is independent of $\delta$ and reported once.

\subsection{Results}
Both the threshold $\theta \in \{0.80, 0.85, 0.90, 0.95\}$ and fusion weight $w \in \{0.0, 0.1, \ldots, 1.0\}$ are jointly selected by grid search on the validation set to maximize $\mathrm{ACC}_{320}$. We then report test-set results using this selected configuration.

\vspace{-2mm}
\begin{table}[h]
\caption{EPD performance for single-task and joint-training system. Best results per block are shown in \textbf{bold}; overall best system are \colorbox{gray!15}{highlighted}.}
\label{tab:all_systems}
\centering
\small
\setlength{\tabcolsep}{3pt}
\renewcommand{\arraystretch}{1.15}
\resizebox{\columnwidth}{!}{%
\begin{tabular}{c|l|c|c|c|cccc}
\hline\hline
\multirow{2}{*}{ID}
& \multirow{2}{*}{Architecture}
& \multirow{2}{*}{Training}
& \multirow{2}{*}{$w$}
& \multirow{2}{*}{$\mathrm{EI}$  $\downarrow$}
& \multicolumn{4}{c}{$\mathrm{ACC}_{\delta}$ $\uparrow$} \\
\cline{6-9}
& & & &
& 160 & 320 & 480 & 640 \\
\hline\hline

1 & Binary
  & Single & --
  & 9.6
  & 78.8 & 83.9 & 84.6 & 84.8 \\

\hline

2 & Duration (REG)
  & Single & --
  & \textbf{8.1}
  & \textbf{79.9} & \textbf{86.4} & \textbf{87.3} & \textbf{87.6} \\

3 & Duration (CLS)
  & Single & --
  & 33.3
  & 63.2 & 66.3 & 66.5 & 66.6 \\

\hline\hline

4 & Bin.\ + Dur.\ (REG)
  & Joint & 0.3
  & 7.7
  & \textbf{80.3} & 85.9 & 86.6 & 86.7 \\

\rowcolor{gray!15}
5 & Bin.\ + Dur.\ (CLS)
  & Joint & 1.0
  & \textbf{5.0}
  & 79.6 & \textbf{86.7} & \textbf{88.1} & \textbf{88.4} \\

\hline\hline
\end{tabular}%
}
\end{table}

\noindent\textbf{Main Results:} 
Table~\ref{tab:all_systems} summarizes the performance of the single-task and joint-training systems. Single REG (ID~2) outperforms the binary baseline (ID~1), achieving lower $\mathrm{EI}$ (8.1\% vs.\ 9.6\%) and higher $\mathrm{ACC}_{\delta}$ at all \(\delta\), indicating that continuous duration prediction provides a stronger endpoint signal. Single CLS alone (ID~3) underperforms Single REG. Joint training with the duration prediction objective (ID~4--5) consistently improves over the binary baseline. Joint REG (ID~4) maintains similar performance to Single REG while improving over the binary baseline in both $\mathrm{EI}$ (7.7\% vs.\ 9.6\%) and $\mathrm{ACC}_{\delta}$ at all
$\delta$. Joint CLS (ID~5) yields the best overall performance ($\mathrm{EI}=5.0\%$, $\mathrm{ACC}_{320}=86.7\%$. Notably, grid search selects \(w=1\) (binary-only inference), suggesting that the primary benefit of Joint CLS comes from the auxiliary supervision.

\vspace{-2mm}
\begin{figure}[h]
    \centering
    \includegraphics[width=0.68\columnwidth]{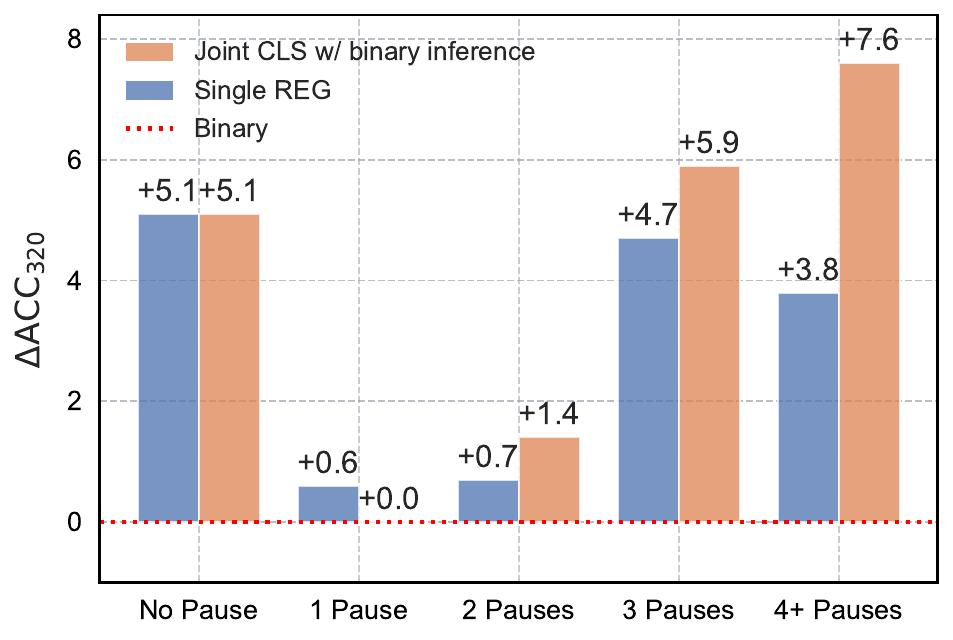}
    \caption{Absolute improvement in $\mathrm{ACC}_{320}$ over the binary baseline for Single REG and Joint CLS with binary inference, broken down by mid-utterance pause count.}
    \label{fig:pause_bar}
\end{figure}

\vspace{-3mm}
\noindent\textbf{Pause-Count Analysis:} Fig.~\ref{fig:pause_bar} shows the absolute \(\mathrm{ACC}_{320}\) gain (over the binary baseline) as a function of pause count for the best single-task system (Single REG) and the best joint-training system (Joint CLS with binary inference), as identified in Table~\ref{tab:all_systems}. The duration-aware systems outperform the binary baseline in every pause count, with gains of +5.1\% in \(\mathrm{ACC}_{320}\) for both systems in the no-pause subset. The gains increase monotonically with more pauses. Single REG improves by +0.6, +0.7, +4.7, and +3.8 \(\mathrm{ACC}_{320}\) points from 1 to 4+ pauses, while Joint CLS with binary inference achieves larger gains at higher pause counts: +0.0, +1.4, +5.9, and +7.6 points. The widening gap with more pauses suggests that duration supervision is particularly beneficial in pause-heavy scenarios.

\begin{table}[h]
\caption{Comparison with acoustic VAD baselines and existing semantic EPD systems under the same streaming protocol in terms of model size, per-chunk latency, and EPD performance.}
\label{tab:comparison}
\centering
\small
\setlength{\tabcolsep}{3pt}
\renewcommand{\arraystretch}{1.15}
\resizebox{\columnwidth}{!}{%
\begin{tabular}{>{\centering\arraybackslash}c|>{\raggedright\arraybackslash}m{0.40\linewidth}|cc|c|cc}
\hline\hline
\multirow{2}{*}{ID}
& \multirow{2}{*}{Model}
& Param.
& Lat. $\downarrow$
& \multirow{2}{*}{$\mathrm{EI}$ $\downarrow$}
& \multicolumn{2}{c}{$\mathrm{ACC}_{\delta}$ $\uparrow$} \\
\cline{3-4}\cline{6-7}
& & (M) & (ms) & & 320 & 640 \\
\hline\hline

\multicolumn{7}{l}{\textit{Acoustic VAD (Silero)}~\cite{silero_vad}}\\
1 & threshold = 320\,ms
& 0.5 & 3
& 60.0
& 33.9 & 39.7 \\

2 & threshold = 480\,ms
& 0.5 & 3
& 45.7
& 10.4 & 52.5 \\

3 & threshold = 640\,ms
& 0.5 & 3
& 27.0
& 5.0 & 61.6 \\

\hline

\multicolumn{7}{l}{\textit{Semantic EPD}}\\
4 & Smart Turn v2~\cite{smartturn}
& 95 & 29
& 59.6
& 32.5 & 35.2 \\

5 & Smart Turn v3.2~\cite{smartturn}
& 8 & 21
& 64.5
& 30.5 & 35.4 \\

6 & Easy Turn~\cite{li2025easyturn}
& 850 & 263
& 31.1
& 60.8 & 67.8 \\

\hline\hline

\multicolumn{7}{l}{\textit{Next-Turn (Proposed)}}\\
\rowcolor{gray!15}
7 & Whisper-large \cite{radford2023whisper} 
& 640 & 152
& \textbf{5.0}
& \textbf{86.7} & \textbf{88.4} \\

8 & Whisper-small \cite{radford2023whisper} 
& 89 & 52
& 10.7
& 81.1 & 83.5 \\

9 & Whisper-tiny \cite{radford2023whisper} 
& 8 & 23
& 12.3
& 73.2 & 78.2 \\

\hline\hline
\end{tabular}%
}
\end{table}

\vspace{-3mm}
\noindent\textbf{Comparison with SOTA Systems:} Table~\ref{tab:comparison} compares our models with open-source models under the same streaming protocol.  The baselines are evaluated as released, without retraining on our corpus or objective, so the comparison reflects realistic end-to-end streaming behavior at deployment rather than a controlled matched-training setting. Acoustic VAD baselines with fixed silence thresholds illustrate the \(\mathrm{EI}\)--\(\mathrm{ACC_{\delta}}\) trade-off: increasing the threshold reduces \(\mathrm{EI}\) from 60.0\% (320\,ms) to 27.0\% (640\,ms), but degrades \(\mathrm{ACC}_{320}\) from 33.9\% to 5.0\%. Semantic EPD baselines (Smart Turn v2/v3.2 and Easy Turn) only process the accumulated speech after a certain silence interval is detected. We use a 160 ms VAD silence threshold for segmentation, which introduces a minimum pre-invocation latency that is included in our evaluation. Despite explicitly modeling turn detection, Smart Turn v2 and v3.2 perform worse than the acoustic baselines, with \(\mathrm{EI}=59.6\%\) and \(64.5\%\), respectively. Easy Turn improves to \(\mathrm{EI}=31.1\%\) with \(\mathrm{ACC}_{320}=60.8\%\), but at the cost of 850\,M parameters and 263\,ms per-chunk latency. Our Whisper-large system achieves the best overall performance (\(\mathrm{EI}=5.0\%\), \(\mathrm{ACC}_{320}=86.7\%\)). Smaller backbones trained with the same procedure remain strong: Whisper-small (89\,M, 52\,ms) attains \(\mathrm{EI}=10.7\%\) and \(\mathrm{ACC}_{320}=81.1\%\), and Whisper-tiny (8\,M, 23\,ms) attains \(\mathrm{EI}=12.3\%\) and \(\mathrm{ACC}_{320}=73.2\%\). Notably, our Whisper-tiny variant matches Smart Turn v3.2 in backbone and parameter count (8\,M) while reducing \(\mathrm{EI}\) by 52.2 absolute points, highlighting the impact of our duration-aware training.

\vspace{-3mm}
\begin{figure}[h]
    \centering
    \includegraphics[width=\columnwidth]{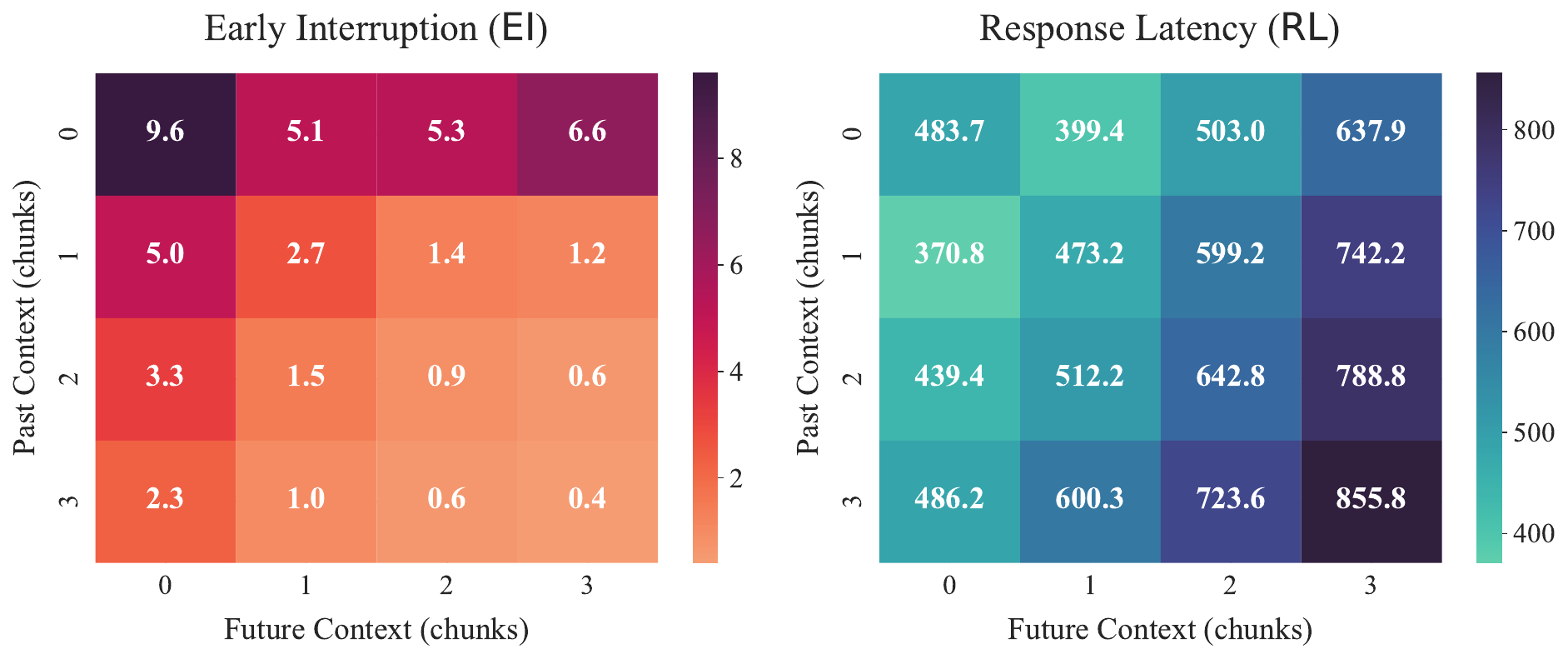}
    \caption{Early interruption ($\mathrm{EI}$, left) and response latency ($\mathrm{RL}$, ms, right) of the overall best-performing system as a function of past ($P$) and future ($F$) context chunks.}
    \label{fig:context}
\vspace{-1mm}
\end{figure}

\vspace{-1mm}
\noindent\textbf{Context Window Analysis:} We investigate the tradeoff between early interruption and response latency when including past ($P$) and future ($F$) context chunks in score post-processing. Incorporating future context can reduce $\mathrm{EI}$ but adds a look-ahead of \(\ell = F \times 160\)\,ms. Here, we define response latency \(\mathrm{RL}\) as the average \(\hat{t}-t_{\text{end}}^{*}\). Fig.~\ref{fig:context} shows \(\mathrm{EI}\) and \(\mathrm{RL}\) as functions of \(P\) and \(F\) for the overall best-performing system (Joint CLS with binary inference). Increasing \(P\) substantially reduces interruptions (at $F=0$, \(\mathrm{EI}\) drops from 9.6\% at $P=0$ to 2.3\% at $P=3$), but can increase \(\mathrm{RL}\) due to temporally smoothed decisions; the lowest \(\mathrm{RL}\) at $F=0$ occurs at $P=1$ (370.8\,ms). Adding future context yields a clear accuracy--latency trade-off: for example, at $P=1$, \(\mathrm{EI}\) decreases from 5.0\% to 1.2\% as \(F\) increases from 0 to 3, while \(\mathrm{RL}\) rises from 370.8\,ms to 742.2\,ms.

\section{Conclusion}
\label{sec:conclusion}
We presented Next-Turn, a duration-aware streaming EPD framework that predicts the time-to-next-speech-onset as a supervision signal derived directly from speech timestamps. Experiments show that the proposed approach outperforms conventional acoustic and recent semantic EPD baselines, achieving a 25.9\% absolute improvement in $\mathrm{ACC}_{320}$ over the strongest baseline. Furthermore, Joint training with the duration-aware objective complements standard binary EPD. Smaller Whisper backbones maintain competitive performance with low latency, supporting low-resource deployment. Future work will extend the framework to multi-turn conversational settings and validate across corpora and languages.

\section{Use of Generative AI Disclosure}
Generative AI tools were used for minor editing and language improvement. All AI-assisted content was reviewed and edited by the authors.




\bibliographystyle{IEEEtran}
\bibliography{refs}

\end{document}